\documentclass[aps, pra, reprint, superscriptaddress]{revtex4-1}
\usepackage{graphicx}
\usepackage{dcolumn}
\usepackage{bm}
\usepackage{amsfonts}
\usepackage{amsmath}
\usepackage{amssymb}
\usepackage{color}

\usepackage[colorlinks=true,citecolor=blue]{hyperref}
\hypersetup{colorlinks=true,citecolor=blue,linkcolor=red,urlcolor=blue}

\def\be{\begin{equation}}
\def\ee{\end{equation}}

\def\kv{{\bf k}}
\def\qv{{\bf q}}

\def\sigmav{{\bm \sigma}}

\begin{document}

\title{Dynamical density response and collective modes of topological-insulator ultra-thin films}

\author{Moslem Mir}
\affiliation{Department of Physics, Institute for Advanced Studies in Basic Sciences (IASBS), Zanjan 45137-66731, Iran}
\affiliation{Faculty of Science, University of Zabol (UOZ), Zabol 98615-538, Iran}
\author{Saeed H. Abedinpour}
\email{abedinpour@iasbs.ac.ir}
\affiliation{Department of Physics, Institute for Advanced Studies in Basic Sciences (IASBS), Zanjan 45137-66731, Iran}
\affiliation{Research Center for Basic Sciences \& Modern Technologies (RBST),  Institute for Advanced Studies in Basic Sciences (IASBS), Zanjan 45137-66731, Iran}
\date{\today}

\begin{abstract}
We analytically calculate the intra- and inter- surface dynamical density-density linear responses of ultra-thin topological insulator films with finite tunneling between their top and bottom surfaces in both metallic and insulating regimes.
Employing the random phase approximation we investigate the dispersions of in-phase and out-of-phase collective density modes of this system in the metallic regime. We find that in contrast to the bilayers of the conventional two-dimensional electron gas, where finite tunneling gaps out the out-of-phase mode, in topological insulator thin films, this mode remains linear at long wavelengths. 
Depending on different system parameters, the velocity of out-of-phase mode can be tuned to be larger or substantially smaller than the Fermi velocity of electrons on the isolated surfaces of the topological insulator.  
Finite tunneling generally reduces the energy of collective modes, making them more confined in space.   
\end{abstract}

\maketitle

\section{Introduction}
Plasmons are collective charge density excitations originating from the long-range nature of the Coulomb interaction. Recently, collective modes of two dimensional Dirac materials have attracted a lot of interest~\cite{Hwang2007,Schutky2013,Pyatkovskiy,L.Ju2011,Grigorenko2012,Stauber2014,C.J.Tabert,Sachdeva2015,AThakur2017,Pietro2013}.
Plasmons in these materials are tunable through gate voltage which controls the carrier concentration, and usually, have higher lifetimes due to the high mobility of host materials. Plasmons have interesting features in topological insulators,  due to the strong spin-orbit coupling~\cite{Zhang T}. They have long propagation lengths and resonance frequencies in the mid-infrared and terahertz spectral regions, that can be tuned via the Fermi energy~\cite{Hsieh}. 
Density oscillations in topological insulators are accompanied by transverse spin oscillations (i.e., spin-plasmons) as a result of spin-orbit coupling~\cite{Efimkin}. 
These peculiar features suggest that the collective modes of topological materials have the potential for novel applications in plasmonics and spintronics. 
Due to the quantum confinement, the bulk band-gap of topological insulator thin film (TITF) is larger than the bulk topological insulator~\cite{Li2015}. This is a useful feature as it provides a larger range of available chemical potentials for the surface states, within the gap of bulk states. 
Nano-scale structures of the topological insulators, such as thin films, multilayers, and nano-ribbons have large surface-to-volume ratios, so the contribution of surface states in their different physical properties are enormously enhanced. 
It has been experimentally observed~\cite{Yi Zhang,Yusuke} that for Bi$_2$Se$_3$ topological insulator thin films when the film thickness becomes less than 6 quintuple layers ($\sim 6 nm$), a finite hybridization between two surfaces opens a gap in the excitation spectrum of surface states. In the following by ultra-thin films, we mean the regime where the tunneling between two surfaces is not negligible. 
Note that a finite gap in the spectrum of surface states of TITF could be induced by other means such as strain~\cite{Mathias2018,Battilomo2019} and magnetic field~\cite{Wang2014, Yang2019}.

Plasmons have been extensively studied in different double layer structures~\cite{DasSarma1998, abedinpour_prl2007, Borghi2009,Gamayun2011, Wang2010, Sensarma2010, Triola2012, Stauber2012, Jin2015,Rosario, Stauber2013,Stauber2017}.  
The collective density modes of topological insulator thin films in the absence of hybridization between their surface states have been investigated recently~\cite{Rosario,Stauber2013,Stauber2017}. Electrons in uncoupled TITF behave like massless Dirac fermions, therefore, their dynamical density response function is similar to the one of graphene~\cite{B. Wunsch}. 
Within the random phase approximation (RPA), it was shown that uncoupled TITF has two collective modes, optical and acoustic modes respectively with the usual $\omega \propto \sqrt{q}$ and $\omega \propto q$ long-wavelength dispersions~\cite{Stauber2013}. 

In this paper, we study the dynamical density-density response and the collective density modes of a topological insulator ultra-thin film, in the regime where the surface electronic states are hybridized due to finite inter-surface tunneling. 
Electrons in tunnel-coupled TITF behave like massive Dirac fermions so their \emph{total} density response at low doping is similar to the density response of massive Dirac fermions~\cite{AThakur2017},  such as in gapped graphene~\cite{Pyatkovskiy} and other buckled honeycomb lattices~\cite{C.J.Tabert}. However, in topological insulator thin films it is possible to probe surface-resolved density responses and therefore two distinct collective modes corresponding to the \emph{in-phase} and \emph{out-of-phase} oscillation of electrons in two surfaces is expected. 
Collective density modes of topological insulator films and double-layer graphene, in the absence of tunneling, have been theoretically investigated by several groups~\cite{Rosario, Stauber2012, Stauber2013, Stauber2014, Stauber2017}. 
Here we look at the effects of hybridization between the surface states of topological insulators in the ultra-thin limit. 
The static density response and screening of this system have been explored by Liu \emph{et al.}~\cite{Weizhe Edward}.

The rest of this paper is organized as follows. In Sec.~\ref{Model_Polarization} we introduce our effective low-energy model Hamiltonian for topological insulator thin film, obtain its intra- and inter-surface density response functions and discuss their behavior in different regimes.
In Sec.~\ref{Plasmons} we discuss the dispersions of collective density modes of TITF, and investigate their long-wavelength behavior. Sec.~\ref{Conclusions} summarizes our main findings.
Finally, the full analytic forms of the dynamical density response functions in different regimes and their asymptotic behavior are presented in Appendix.~\ref{app:TITF_Response}.

\section{\label{Model_Polarization} Model Hamiltonian and linear density-density responses}
The effective low-energy single-particle Hamiltonian of a TITF is given by~\cite{Pouyan Ghaemi,Zyuzin,Sergey,Weizhe Edward,Fariborz}
\be\label{Hamiltonian}
{\hat {\cal H}}_{\kv}=\hbar v_{\rm F}\tau_{z}\otimes[\sigmav\cdot(\kv\times \hat{z})]+t\tau_{x}\otimes \sigma_0,
\ee
where $v_{\rm F}$ is the Fermi velocity of the surface states, $\tau$ and $\sigma$ are the Pauli matrices acting in the layer and real spin spaces, respectively and their zero components are $2\times 2$ identity matrices in the corresponding space, $t$ is the tunneling between top and bottom surfaces and $\hat{z}$ is a unit vector in the direction perpendicular to the surfaces.
Note that the first term on the right-hand-side of Eq.~\eqref{Hamiltonian} describes the Dirac fermions on two surfaces of a topological insulator while the second term is the hybridization of two surfaces due to the tunneling of electrons in ultra-thin films. This tunneling is responsible for the gap opening in the dispersion of surface states. 
We should note that the surface states of a tunnel coupled topological insulator thin film may acquire a nontrivial Chern number showing quantum spin Hall effect~\cite{Yi Zhang, Hai-Zhou2010,Wen2010,asmar_prb2018}.
The inclusion of at least quadratic terms in momentum in the Hamiltonian of TITF is necessary to capture such topological effects. 
Our main focus in this work, however, is on the collective modes of the surface electrons in the regime where the Fermi energy resides above the tunneling induced gap of the surface states. 
The edge states are not expected to have a significant effect on the surface collective modes. 
Therefore, here we resort to the minimal effective model given by Eq.~\eqref{Hamiltonian} which captures the main low energy features of the surface states of a TITF and at the same time is simple enough to make the analytical treatment of dynamical responses and collective modes feasible.

Diagonalizing the effective $4\times4$ Hamiltonian \eqref{Hamiltonian}, one obtains the dispersions of valence and conduction bands
\be\label{eigenenergy}
 \varepsilon_{k\lambda}=\lambda \sqrt{\hbar^2 v^{2}_{F} k^2+t^2},
\ee
where $\lambda=-1 (+1)$ refers to two spin-degenerate valence (conduction) bands. 
The corresponding normalized eigenstates in the 
$\psi=(\psi_{1,\uparrow},\psi_{1,\downarrow}, \psi_{2,\uparrow}, \psi_{2,\downarrow})^{\cal T}$ 
basis, where two indices specify the layer and spin orientation of electrons, read~\cite{Weizhe Edward}
\be\label{eq:eigenstates}  
\begin{split}
\Psi_{\kv\lambda}^{(1)}&=\frac{1}{\sqrt{2}}
    \left(\lambda,-i\cos \alpha_k e^{i\theta_{k}},\sin\alpha_k,0\right)^{\cal T}, \\
    \Psi_{\kv \lambda}^{(2)}&=\frac{1}{\sqrt{2}}
    \left(0,\sin\alpha_k,-i\cos\alpha_k e^{-i\theta_{k}},\lambda\right)^{\cal T},
\end{split}
\ee
with ${\cal T}$ referring to the transpose of a vector, and $\alpha_k \equiv \tan^{-1} (t/\hbar v_{\rm F} k)$ and $\theta_{ k}\equiv\tan^{-1}(k_y/k_x)$, are defined for notational convenience.

The density fluctuations $\delta \rho_l(q,\omega)$ induced in the top ($l=1$) or bottom ($l=2$) surface, in the linear response regime are given by~\cite{Vignale_Book}
\be
\delta \rho_l(q,\omega)=\sum_{l'}\chi_{ll'}(q,\omega) V^{\rm ext}_{l'}(q,\omega),
\ee
where $\chi_{ll'}(q,\omega)$ is the surface-resolved linear density-density response function and $V^{\rm ext}_{l'}(q,\omega)$ is the external potential applied to surface $l'$. 
Within the random phase approximation the interacting linear density-density response function could be written in the matrix form as
\be\label{RPA}
\chi^{\rm RPA}(q,\omega)=[1-\Pi(q,\omega)V(q)]^{-1}\Pi(q,\omega),
\ee
where
\begin{equation}
\Pi(q,\omega)=\begin{pmatrix}
 \Pi_{11}(q,\omega)    &  \Pi_{12}(q,\omega)\\  
 \Pi_{21}(q,\omega)    &  \Pi_{22} (q,\omega)
 \end{pmatrix},
\end{equation}
and 
\begin{equation}\label{eq:V-matrix}
V(q)=\begin{pmatrix} V_{11}(q) & V_{12}(q)\\
V_{21}(q) &V_{22}(q)\end{pmatrix},
\end{equation}
are the matrices of non-interacting density-density response function and Coulomb interaction~\cite{Jin2015}, respectively.
\begin{figure}
\centering
\includegraphics[width=1.1\linewidth]{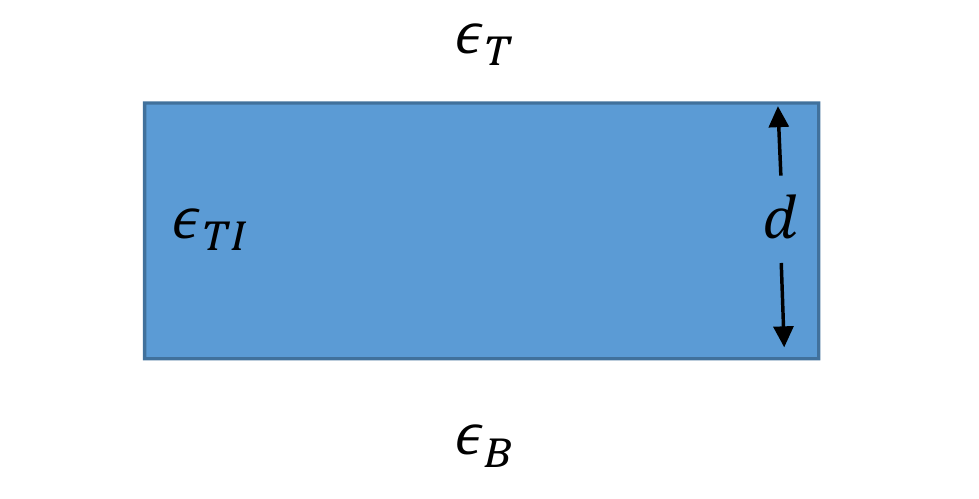}
\caption{Schematic picture of an ultra-thin film of topological insulator with thickness $d$ and dielectric constant $\epsilon_{\rm TI}$, surrounded by different mediums on top (with the dielectric constant $\epsilon_{\rm T}$) and bottom (with the dielectric constant $\epsilon_{\rm B}$).
\label{fig:slab}}
\end{figure}
If we assume that a thin slab of topological insulator with the thickness of $d$ and the dielectric constant of $\epsilon_{\rm TI}$ is sandwiched between different dielectric media, on the top and bottom with the dielectric constants $\epsilon_{\rm T}$ and $\epsilon_{\rm B}$, respectively (see, Fig.~\ref{fig:slab}), the Coulomb interaction between electrons in the top surface would be given by~\cite{Rosario}
\be\label{intralayer_coulomb}
V_{11}(q)= \frac{\epsilon_{\rm TI}\cosh(qd)+\epsilon_{\rm B}\sinh(qd)}{D(q)} v_q,
\ee
and the inter-surface Coulomb interaction reads
\be\label{interlayer_coulomb}
V_{12}(q)=V_{21}(q)=\frac{\epsilon_{\rm TI}}{D(q)}v_{q},
\ee
and the Coulomb interaction in the bottom surface $V_{22}(q)$ is obtained from interchanging $\epsilon_{\rm T} \leftrightarrow \epsilon_{\rm B}$ in Eq.~\eqref{intralayer_coulomb}.
Here, $v_q=2\pi e^2/q$ and
\be
D(q)=\frac{\epsilon_{\rm TI}(\epsilon_{\rm T}+\epsilon_{\rm B})\cosh(qd)+(\epsilon_{\rm T} \epsilon_{\rm B}+\epsilon_{\rm TI}^{2})\sinh(qd)}{2}.
\ee
Note that, in the uniform i.e., $\epsilon_{\rm TI}=\epsilon_{\rm T}=\epsilon_{\rm B}=1$ limit, we recover the familiar expressions $V_{11}(q)=V_{22}(q)=v_q$, and $V_{12}(q)=v_q e^{-qd}$, for the intra-surface and inter-surface interactions, respectively.

The components of the surface-resolved non-interacting density response functions are given by
\be\label{chi_0} 
\Pi_{ll'}(q,\omega)=
\frac{1}{S}\sum_{\kv,\lambda,\lambda'}
\frac{F^{ll'}_{\lambda\lambda'}(\kv,\kv')\left[f(\varepsilon_{k\lambda})-f(\varepsilon_{k'\lambda'})\right] }{\hbar \omega + \varepsilon_{k\lambda}-\varepsilon_{k'\lambda'}+i 0^+ },
\ee
where $S$ is the surface area, $\kv' \equiv\kv+\qv$, $f(\varepsilon)$ is the Fermi distribution function, and the form-factors are
\be\label{form}
F^{ll'}_{\lambda\lambda'}(\kv,\kv')=\sum_{i,j} 
\langle\Psi^{(i)}_{\kv\lambda} | \rho^{l}|\Psi^{(j)}_{\kv'\lambda'} \rangle \langle \Psi^{(j)}_{\kv'\lambda'} \vert \rho^{l'} \vert \Psi^{(i)}_{\kv\lambda} \rangle,
\ee
with the surface-resolved density operators of TITF defined as $\rho^l \equiv [\tau^0-(-1)^l \tau^z]\otimes \sigma^0/2$.
It is straightforward to show that
\be\label{f11}
F^{ll}_{\lambda\lambda'}(\kv,\kv')
=\frac{1}{2}\left(1+\lambda\lambda'\frac{ \hbar^2 v^{2}_{\rm F}\kv\cdot\kv'}{\varepsilon_{k}\varepsilon_{k'}}\right),
\ee
and
\be\label{f12}
F^{12}_{\lambda\lambda'}(\kv,\kv')=F^{21}_{\lambda\lambda'}(\kv,\kv')
=\frac{1}{2}\frac{\lambda\lambda' t^2}{\varepsilon_{k}\varepsilon_{k'}}.
\ee
Here, $\varepsilon_{k}\equiv\vert \varepsilon_{k\lambda} \vert$ is introduced for notational convenience. 
We should note that the total density response function
\be 
\begin{split}
\Pi(q,\omega)&=\sum_{l,l'}\Pi_{ll'}(q,\omega)\\
&=\frac{1}{S}\sum_{\kv, \lambda,\lambda'}\frac{f(\varepsilon_{k\lambda})-f(\varepsilon_{k'\lambda'})}{\hbar\omega+\varepsilon_{k\lambda}-\varepsilon_{k'\lambda'}+i0^+} \\
&~~~~~~~~~
\times
\left(1+\lambda \lambda'\frac{\hbar^2 v^{2}_{\rm F}\kv\cdot\kv'+t^2}{\varepsilon_{k}\varepsilon_{k'}}\right),
\end{split}
\ee
apart from a trivial degeneracy factor, is identical to the dynamical density response function of two-dimensional massive Dirac fermions~\cite{Pyatkovskiy, AThakur2017, Mazloom_PRB2016}. 
It is possible to find analytic expressions for the real and imaginary parts of the layer-resolved density-density response functions for arbitrary frequency and wave-vector (c.f. Appendix~\ref{app:TITF_Response}).  
As the total density response function $\Pi(q,\omega)$ is analytically known~\cite{Pyatkovskiy}, we have simply presented the analytic expressions for the inter-surface component of the density response function $\Pi_{12}(q,\omega)$ in Appendix~\ref{app:TITF_Response}. The intra-surface component of the density response function could be readily obtained from $\Pi_{11}(q,\omega)=\Pi(q,\omega)/2-\Pi_{12}(q,\omega)$. 
\begin{figure}
\centering
\includegraphics[width=1.1\linewidth]{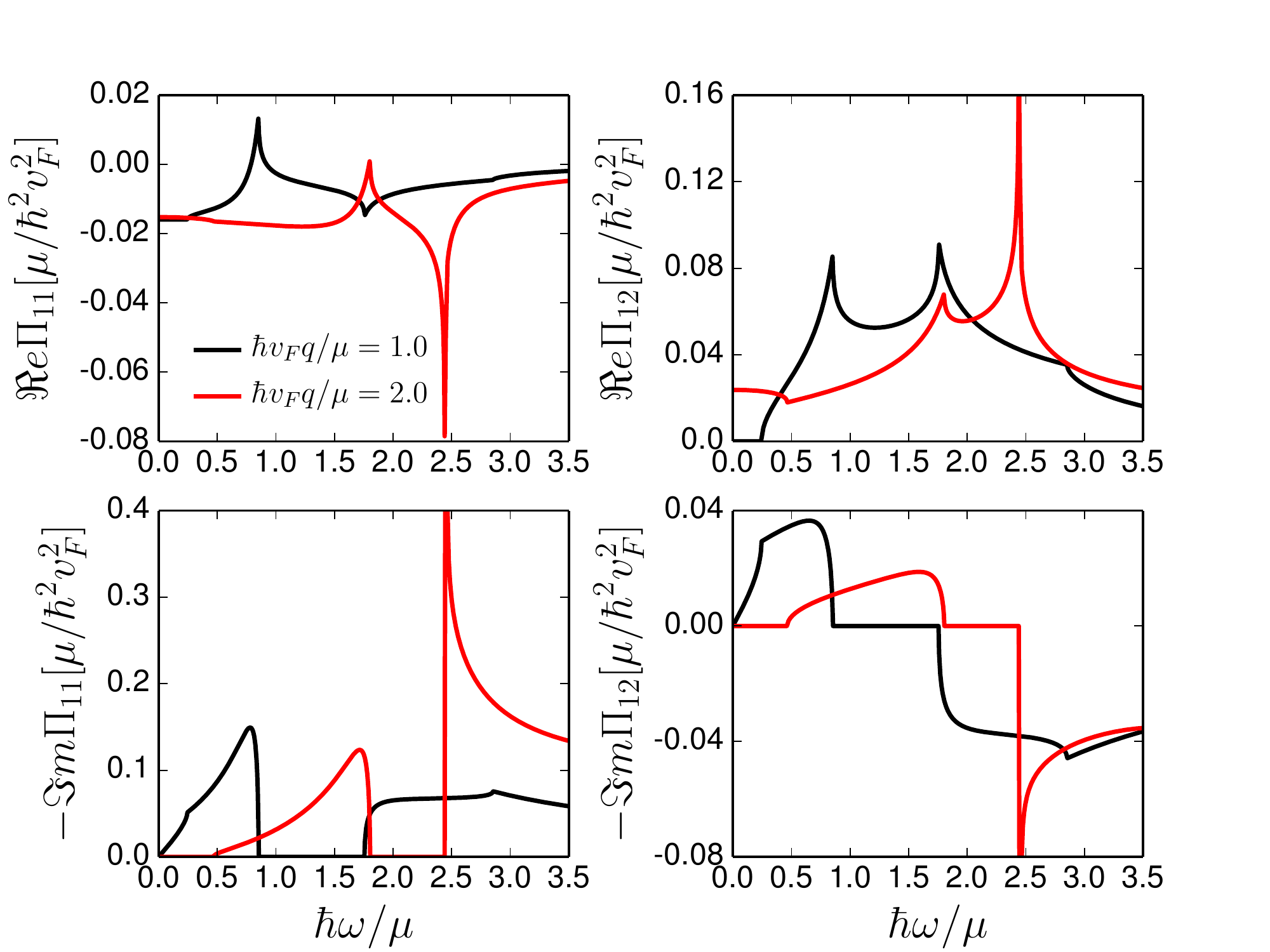}
\caption{The real (top) and imaginary (bottom) parts of intra-surface (left) and inter-surface (right) components of the non-interacting density-density response function of a topological insulator ultra-thin film versus frequency for two fixed values of 
the wave vector. The tunneling between two surfaces is set equal to $t=0.7 \,\mu$, where $\mu$ is the chemical potential.}
\label{fig:chi_q}
\end{figure}
The behavior of dynamical density response function versus frequency at two representative wave-vectors is illustrated in Fig.~\ref{fig:chi_q}.
The imaginary parts of both intra-surface and inter-surface density responses are non-zero inside the intra-band and inter-band electron-hole excitation continuum, where the dissipation due to single particle excitations is allowed.
Sharp changes in both real and imaginary parts of the response functions characterize the boundaries of the electron-hole continuum in the $(\qv,\omega)$-plane (c.f. Fig.~\ref{fig:SPE} for an illustration of different regions).

\subsection{The static limit}
\begin{figure}
\centering
\includegraphics[width=1.1\linewidth]{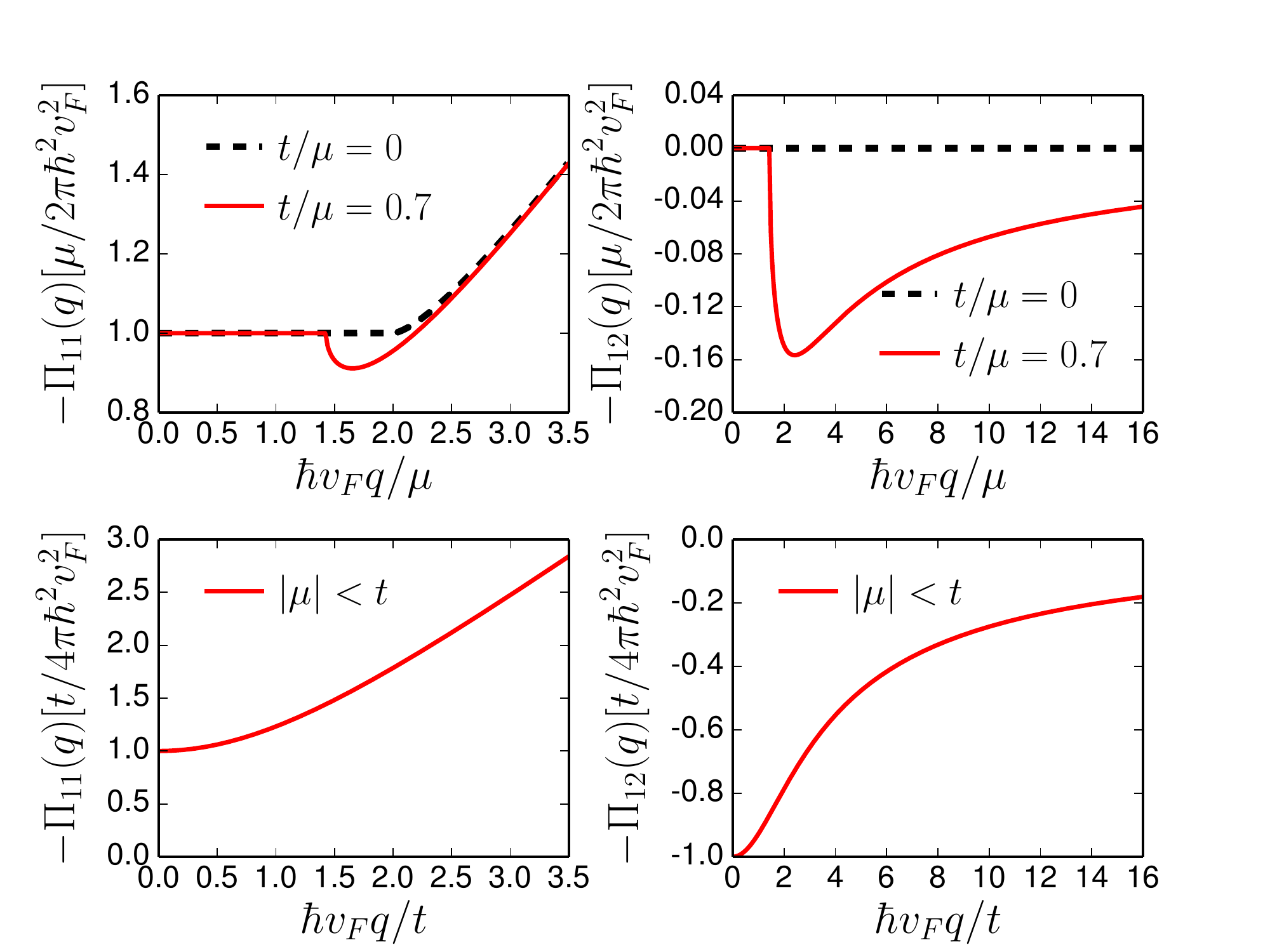}
\caption{Top: the intra-surface (left) and inter-surface (right) components of the non-interacting static density-density response function of an electron doped topological insulator ultra-thin film versus wave vector for $t=0.7\,\mu$ (solid red curves). Results in the vanishing tunneling limit (dashed black curves) are also presented for comparison.
Bottom: same as the top panels but for the case when the chemical potential lies inside the band gap i.e., $|\mu| < t$.
\label{fig:static_screening}}
\end{figure}
The static density-density response functions are obtained from the $\omega \to 0$ limit of the dynamical responses. 
In the electron doped metallic regime i.e., $\mu> t$, where $\mu$ is the chemical potential of system, we find 
\be\label{static_intra}
\begin{split}
&\Pi_{11}(q)=-\frac{\mu}{2\pi \hbar^2 v^2_{\rm F}} 
 \Bigg\{1-\Theta(q-2k_{\rm F})   \\
& \times  
\left[\frac{\sqrt{q^2-4k^2_{\rm F}}}{2q}-\frac{\hbar v_{\rm F}q}{4\mu}\arctan\left(\frac{\hbar v_{F}\sqrt{q^2-4k^2_{\rm F}}}{2\mu}\right)\right]\Bigg\},
\end{split}
\ee
and
\be\label{static_inter}
\begin{split}
\Pi_{12}(q)=\frac{t^2 \Theta(q-2k_{\rm F})}{2\pi \hbar^3 v^3_{\rm F}q}
\arctan\left(\frac{\hbar v_{\rm F}\sqrt{q^2-4k^2_{\rm F}}}{2\mu}\right),
\end{split}
\ee
where $k_{\rm F}=\sqrt{\mu^2-t^2}/(\hbar v_{\rm F})$ is the Fermi wave vector in the conduction band and $\Theta(x)$ is the Heaviside step function.
If the chemical potential lies inside the band gap i.e., $|\mu|<t$, the system at zero temperature behaves like an insulator and we get
\be\label{Re_chi0_11_q}
 \Pi_{11}(q)=-\frac{t}{4\pi \hbar^2 v_{\rm F}^2}
 -\frac{q}{16\pi \hbar v_{\rm F}} \arccos\left(\frac{4t^2-\hbar^2 v_{F}^2q^2}{4t^2+\hbar^2 v_{F}^2q^2}\right),
\ee
and
\be\label{Re_chi0_12_q}
\Pi_{12}(q)=
\frac{t^2}{4\pi\hbar^3 v_{\rm F}^3 q} \arccos\left(\frac{4t^2-\hbar^2 v_{F}^2q^2}{4t^2+\hbar^2 v_{F}^2q^2}\right),
\ee
which are clearly independent of the chemical potential.
The behavior of the inter-surface and intra-surface components of the static non-interacting density-density response function of a TITF  in the static regime is illustrated in Fig.~\ref{fig:static_screening}. 
In the metallic regime (Fig.~\ref{fig:static_screening}, top panels), the static inter-surface response vanishes for $q<2 k_{\rm F}$, where $k_{\rm F}$ is the Fermi wave vector. The intra-surface response is constant in the same regime, typical behavior of two dimensional systems~\cite{Vignale_Book}. 
Restoration of the backscattering gives rise to the suppression of both intra-surface and inter-surface static density-density response functions for $q>2 k_{\rm F}$~\cite{Weizhe Edward}. At large wave vectors, the inter-surface response vanishes but the absolute value of the intra-surface response linearly increases with $q$. This linear dependence of the static density response function on the wave vector is similar to other two-dimensional Dirac materials and originates from the interband transitions ~\cite{Hwang2007}. 
Combined with the $1/q$ behavior of the bare Coulomb interaction, this gives rise to the enhancement of the effective dielectric constant of the medium. 
When the chemical potential lies inside the bandgap (Fig.~\ref{fig:static_screening}, bottom panels), the total density response vanishes in the long-wavelength limit, as expected for an insulator, however, the surface resolved responses are finite.
We should note that these behaviors have been already investigated for the inter-band and intra-band components of the static density-density response function of TITF by Liu \emph{et al.}~ \cite{Weizhe Edward}.

\subsection{The dynamical responses in the $q=0$ limit}
In the vanishing wave vector limit, only the inter-band transitions contribute to the dynamical density-density response functions and for an electron doped metallic system (i.e., $\mu>t$) we obtain
\be\label{Re_chi_dynamic}
\Re e\, \Pi_{11}(0,\omega)=-\frac{t^2}{4\pi \hbar^3  v_{\rm F}^2  \omega}
\ln \left(\left|\frac{2\mu+\hbar\omega}{2\mu-\hbar \omega }\right|\right),
\ee
and
\be\label{Im_chi_dynamic}
\Im m\, \Pi_{11}(0,\omega)=-\frac{t^2}{4 \hbar^3 v_{\rm F}^2\omega}
\Theta ( \hbar \omega- 2\mu),
\ee
for the real and imaginary parts of the intra-surface density-density response functions, respectively. 
Also note that we have $\Pi_{12}(0,\omega)=-\Pi_{11}(0,\omega)$, and the total dynamical density-density response function vanishes for $q=0$.
In the $|\mu|<t$ limit, replacing the chemical potential $\mu$ in Eqs.~\eqref{Re_chi_dynamic} and \eqref{Im_chi_dynamic} with the tunneling $t$, we find the corresponding dynamical density-density response functions in the insulating regime.

\section{\label{Plasmons} Collective modes}
Collective density oscillations could be obtained from the poles of the interacting density response function~\eqref{RPA},  which is equivalent to the solutions of the following equation
\be\label{mode}
\left[1-\lambda_{+}(q,\omega)\right]\left[1-\lambda_{-}(q,\omega)\right]=0,
\ee
with 
\be
\begin{split}
\lambda_{\pm}=&\frac{V_{+}\Pi_{+}+V_{-}\Pi_{-}}{2}\\
&\pm\sqrt{\left(\frac{V_{+}\Pi_{+}-V_{-}\Pi_{-}}{2}\right)^2+V_{a}^{2}\Pi_{+}\Pi_{-}} ~,
\end{split}
\ee
where $\Pi_{\pm}=\Pi_{11} \pm \Pi_{12}$, $V_{\pm}=(V_{11}+V_{22})/2 \pm V_{12}$, and $V_{a}=(V_{11}-V_{22})/2$.
Note that for $\epsilon_{\rm B}=\epsilon_{\rm T}$, we have $V_{11}=V_{22}$, then Eq.~\eqref{mode} simplifies to
\be\label{mode_1}
\left[1-V_+(q) \Pi_+(q,\omega)\right]\left[1-V_-(q) \Pi_-(q,\omega)\right]=0,
\ee
and the symmetric and asymmetric modes become totally decoupled.
The solutions of Eq.~\eqref{mode} result in two collective density mode branches $\omega_+(q)$ and $\omega_-(q)$, corresponding respectively to the optical and acoustic density modes.
Un-damped collective modes occur in the regions of the $(\qv,\omega)$-plane where the imaginary parts of the density response functions are zero, i.e., outside the electron-hole continuum (EHC). The analytic expressions for  the non-interacting density response functions $\Pi_{ll'}(q,\omega)$ are provided in Appendix~\ref{app:TITF_Response}. The imaginary parts of the density response functions are non-zero only in regions 3A, 4A, 1B and 5B [see, Eq.~\eqref{Regions} and Fig.~\ref{fig:SPE} for the definitions]. Note that at zero temperature no collective density oscillation could be excited in the insulating regime (i.e., $|\mu|<t$). Therefore, in the following, we discuss the dispersions of collective modes in the metallic regime.\\

\subsection{Analytic results in the long-wavelength limit}
We begin with the presentation of our analytic results for the dispersions of collective modes in the long wavelength, i.e., $q\to 0$ limit. 
For the optical mode $\omega_+(q)$ in the long wavelength limit, upon substituting the 
dynamical long wavelength limit behaviors of density response functions from Eqs.~\eqref{chi_+11_q_go_zero} and~\eqref{chi_+12_q_go_zero}, together with the long wavelength behaviors of the Coulomb interactions from Eqs.~\eqref{V_11_q_go_to_zero} and \eqref{V_12_q_go_to_zero}, into the first square bracket on the left-hand-side of Eq.~\eqref{mode}, we obtain
\be\label{plasmon_small_q}
\frac{\hbar\omega_{+}(q\to 0)}{\mu}
\simeq
\sqrt{1-\bar{t}^2}\sqrt{\frac{2 \alpha_{ee}\hbar v_{\rm F} q}{(\epsilon_{\rm T}+\epsilon_{\rm B})\mu }}+{\cal O}(q^{3/2}),
\ee
which has the expected $\sqrt{q}$ dependance of the plasmon dispersion in two dimensions~\cite{Vignale_Book}. 
Here, ${\bar t}=t/\mu <1$ is the dimensionless tunneling parameter, and $\alpha_{ee}=e^2/(\hbar v_{\rm F})$ is the dimensionless coupling constant of massless Dirac electrons.
This expression is valid for small wave vectors i.e., $q d \ll (\epsilon_{\rm T}+\epsilon_{\rm B})/\epsilon_{\rm TI}$, and it is interesting to note that the long-wavelength behavior of the optical mode is not sensitive to the dielectric constant of the topological insulator. 
For intermediate values of $q$, it is possible to find the leading order quantum correction to the mode dispersion~\cite{Sensarma2010, Stauber2013}
\be\label{plasmon_larger_q}
\frac{\hbar\omega_{+}(q\to 0)}{\mu}
\simeq
\sqrt{1-\bar{t}^2}\sqrt{\frac{2\alpha_{ee}\hbar v_{\rm F} q}{(\epsilon_{\rm T}+\epsilon_{\rm B})\mu }\left(1+\frac{qd\epsilon_{\rm TI}}{\epsilon_{\rm T}+\epsilon_{\rm B}}\right)^{-1}},
\ee
which is valid for $q d\ll 1$.  

\begin{figure}
\centering
\includegraphics[width=1.1\linewidth]{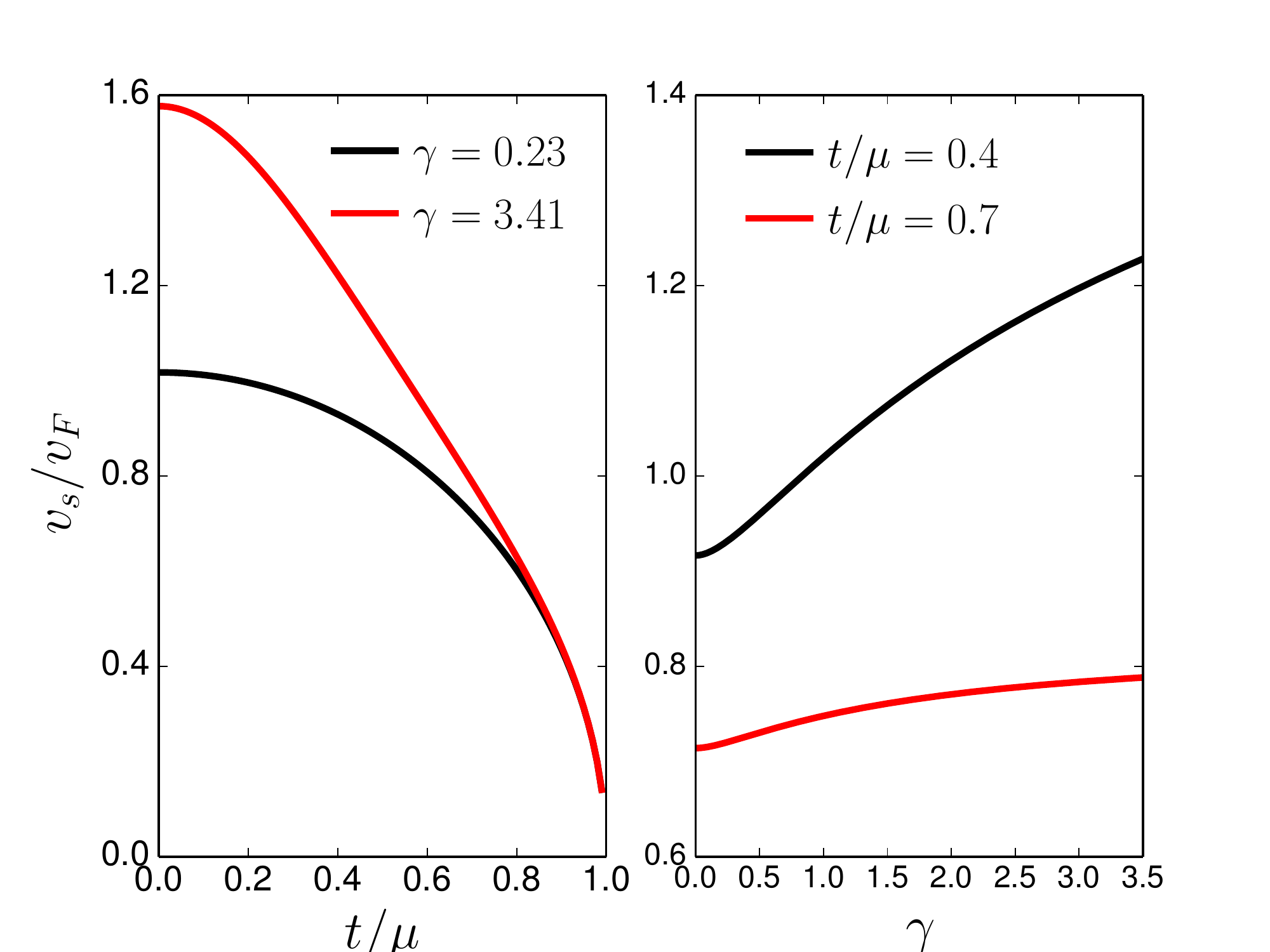}
\caption{The sound velocity $v_s$ (in the units of Fermi velocity $v_{\rm F}$) versus interlayer tunneling parameter for fixed values of $\gamma$ (left)  and versus coupling parameter $\gamma$ for fixed values of tunneling (right).
}\label{fig:acoustic_long}
\end{figure}

To obtain the acoustic mode $\omega_-(q)$ in the $q\to 0$ limit, the long wavelength behaviors of the density response functions in the acoustic limit from Eqs.~\eqref{chi_-11_q_go_zero} and~\eqref{chi_-12_q_go_zero}, together with the long wavelength limits of the Coulomb interactions should be substituted into the second square bracket on the left-hand-side of Eq.~\eqref{mode} to give
\be\label{acoustic_mode}
\omega_{-}(q\to 0) \simeq v_s q
= \frac{\sqrt{1-\bar{t}^2} }{g(\gamma,\bar{t})} v_{\rm F} q ,
\ee
with
\be\label{eq:g}
g(\gamma,\bar{t})={\sqrt{1-\left(\frac{1-{\bar t}^2}{1+\sqrt{1-{\bar t}^2}/\gamma}\right)^2}},
\ee
and $\gamma=\alpha_{ee} k_{\rm F} d/\epsilon_{\rm TI}$.
As expected, the out-of-phase oscillation of electrons in two surfaces is influenced by the dielectric constant of the TITF through $\gamma$.
Note that the border between $4A$ and $2A$ regions (c.f., Fig.~\ref{fig:SPE}) at the long wavelength limit is given by $\omega= \sqrt{1-\bar{t}^2}v_{\rm F} q $. As $g(\gamma,\bar{t})$ is always smaller than one, we have $v_s > \sqrt{1-\bar{t}^2}v_{\rm F}$, and the acoustic mode is always undamped at long wavelengths.
For $v_s > v_{\rm F}$ this mode lies in region $5B$, whereas for $\sqrt{1-\bar{t}^2}v_{\rm F} < v_s < v_{\rm F}$ it resides in region $4A$. 
As we have illustrated in Fig.~\ref{fig:acoustic_long}, depending on different system parameters, the velocity of acoustic mode $v_s$ can be larger or substantially smaller than the Fermi velocity of isolated surface states $v_{\rm F}$. 

In the absence of inter-surface hybridization i.e., $t=0$, we have $\mu=\hbar v_{\rm F} k_{\rm F}$, and Eqs.~\eqref{plasmon_small_q},~\eqref{plasmon_larger_q} and~\eqref{acoustic_mode} all reduce to the standard results obtained for double layer graphene and topological insulator thin films in the absence of tunneling~\cite{Stauber2013, Rosario}.

\subsection{\label{Numerical_result}Numerical results on the dispersions of collective modes}
The dispersions of the collective modes for arbitrary wave vectors could be obtained from the numerical solution of Eq.~\eqref{mode}. 
The system parameters we have used in our numerical calculations are
$t \approx 126$ meV and $v_{\rm F} \approx 4.71 \times10^5$ ms$^{-1}$, which results in $\hbar v_{\rm F} \approx 3.1$ eV$\AA$ and $\alpha_{\rm ee} \approx 4.65$. 
These values correspond to an ultra-thin film of Bi$_2$Se$_3$ with the thickness of two quintuple layers $d\approx20\AA$, whose electronic structure has been explored experimentally through the angel resolved photo-emission spectroscopy~\cite{Yi Zhang}.  
Unless otherwise stated, we are going to use $\epsilon_{\rm TI}=30$ for the dielectric constant of the Bi$_2$Se$_3$ thin film, which is smaller than its bulk value (i.e., $\epsilon_{\rm Bulk} \approx 100$)~\cite{autore}. We also take $\epsilon_{\rm T}=1$ and $\epsilon_{B}=10$, corresponding to the dielectric constants of air and sapphire substrate $\rm Al_{2}O_{3}$, respectively~\cite{Stauber2013}.

\begin{figure} 
\centering
\begin{tabular}{c}
\includegraphics[width=\linewidth]{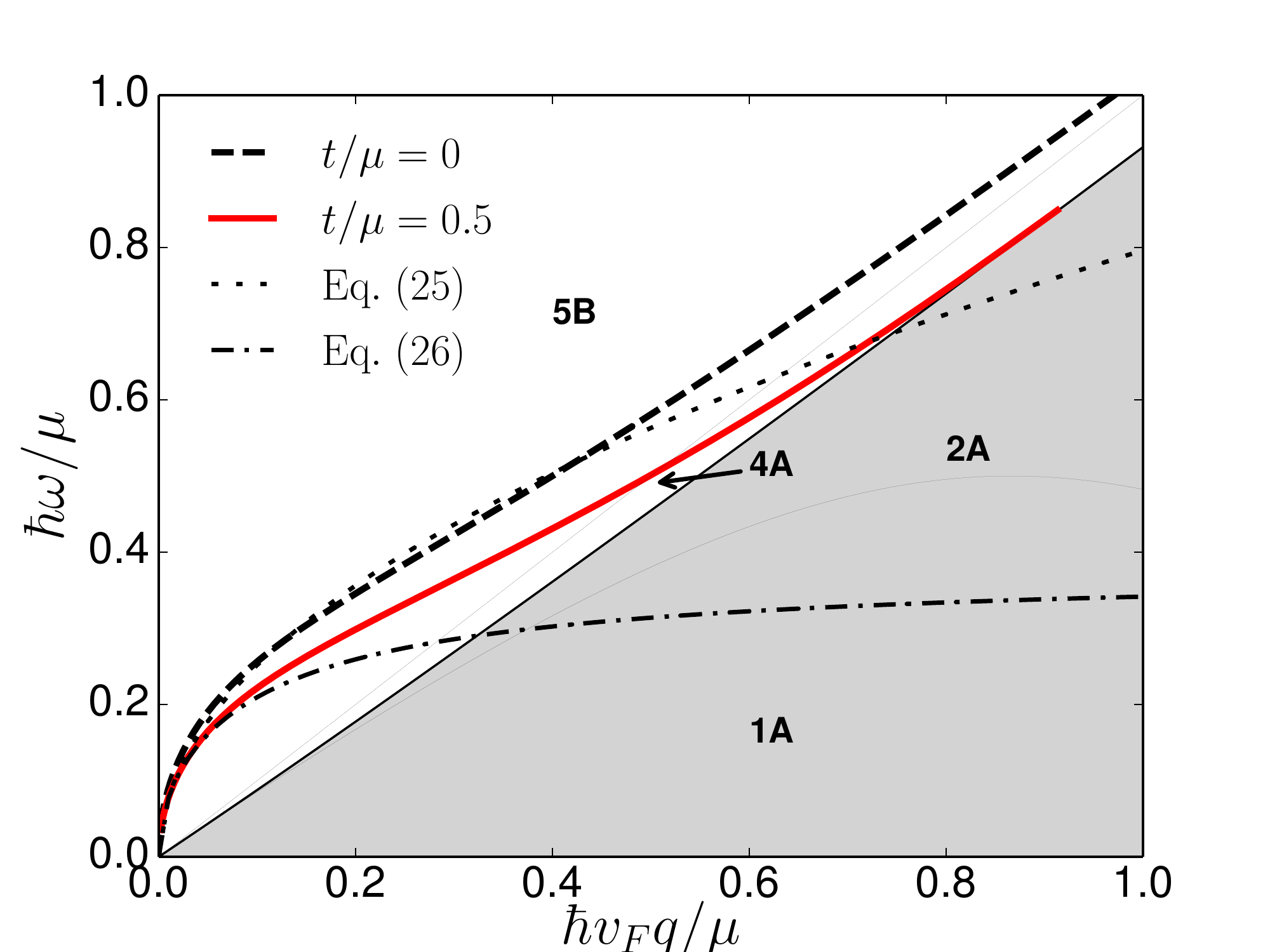}
\end{tabular}
\caption{The full dispersion of the optical mode (solid red) of a topological insulator ultra-thin film is compared to its analytic long wavelength solutions obtained from Eqs.~\eqref{plasmon_small_q} (dotted black) and~\eqref{plasmon_larger_q} (dash-dotted black). 
The chemical potential $\mu=252$ meV and the tunneling $t = 126$ meV has been used here.
The full dispersion in the absence of tunneling (dashed black) is also plotted for comparison.
For the dielectric constants we have used $\epsilon_{\rm TI}=30$, $\epsilon_{\rm T}=1$, and $\epsilon_{\rm B}=10$.
Thin solid black lines separate different regions in the $(q,\omega)$-plane, as introduced in Appendix~\ref{app:TITF_Response} and the gray area refers to the electron-hole continuum where the imaginary part of the non-interacting density response functions are non-zero.
\label{fig:optical}}
\end{figure}

In Fig.~\ref{fig:optical} the dispersion of optical collective mode $\omega_+(q)$ is illustrated. The full numerical results are compared with the analytical expressions in Eqs.~\eqref{plasmon_small_q} and \eqref{plasmon_larger_q}. 
Note that the dispersion of optical mode from Eq.~\eqref{plasmon_larger_q} gives a plateau at large wave vectors, while such a feature is absent in the full numerical dispersion.
It is also evident that the finite tunneling reduces the energy of the symmetric mode. 

\begin{figure}
\centering
\begin{tabular}{c}
\includegraphics[width=\linewidth]{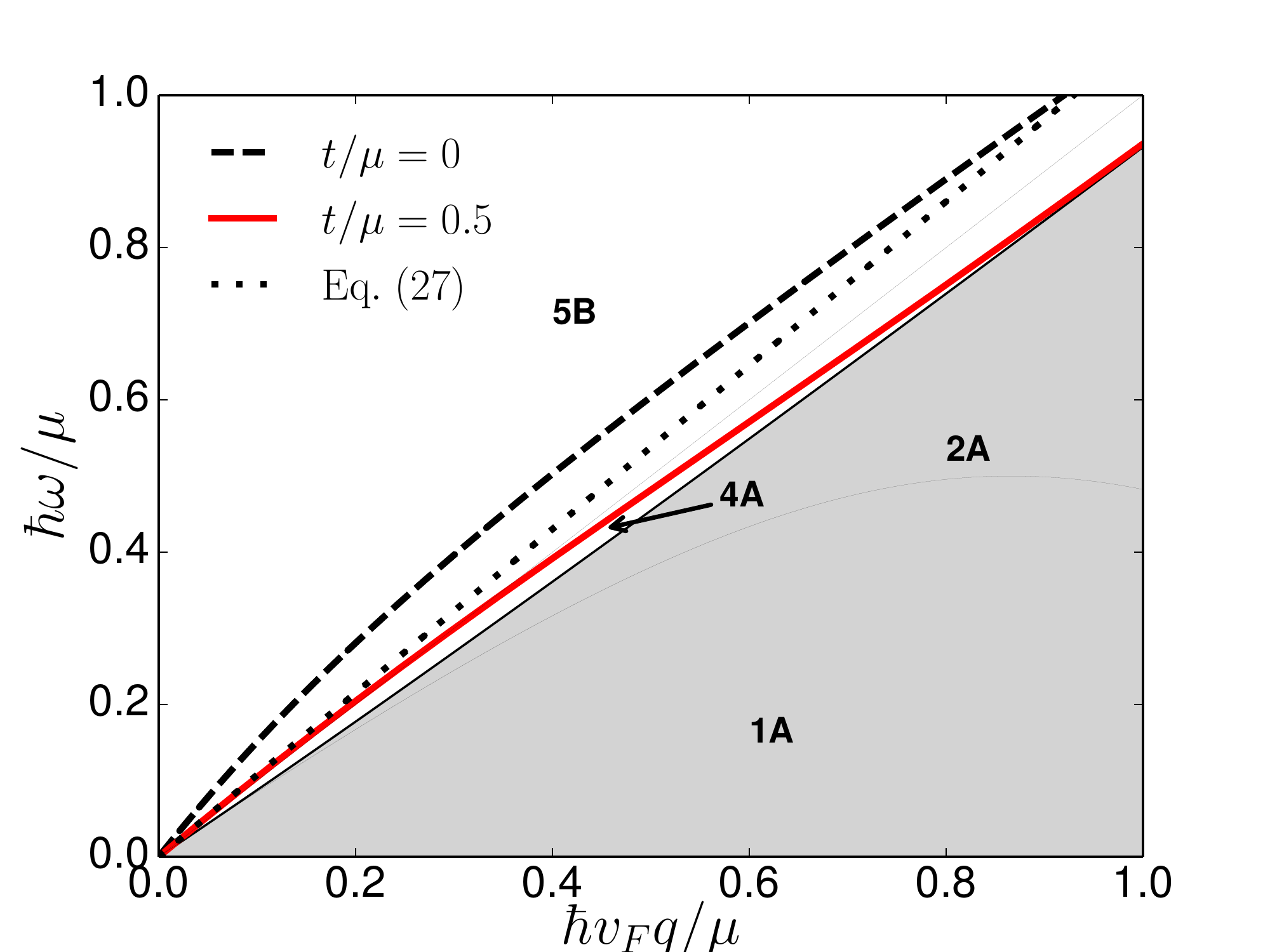}
\end{tabular}
\caption{Full dispersion of the acoustic collective mode (solid red) of a topological insulator ultra-thin film is compared to its analytic long wavelength solution from Eq.~\eqref{acoustic_mode} (dotted black).
 The chemical potential $\mu=252$ meV has been used here together with $t = 126$ meV.
The full dispersion of the acoustic mode in the absence of tunneling (dashed black) is also plotted for comparison.
For the dielectric constants we have used $\epsilon_{\rm TI}=2$, $\epsilon_{\rm T}=1$, and $\epsilon_{\rm B}=10$.
Thin solid black lines separate different regions in the $(q,\omega)$-plane, as introduced in Appendix~\ref{app:TITF_Response} and the gray area refers to the electron-hole continuum where the imaginary part of the non-interacting density response functions are non-zero.\label{fig:acoustic}}
\end{figure}

The full dispersion of the acoustic mode is shown in Fig.~\ref{fig:acoustic}. Here, for better visibility, we have used a much smaller value for the dielectric constant of the topological insulator thin film (i.e., $\epsilon_{\rm TI}=2$) as the mode dispersion for $\epsilon_{\rm TI}=30$ lies too close to the boundary of EHC. Again, it is evident that finite tunneling lowers the energy of the collective mode.
Notice that in the presence of a finite tunneling between two surfaces, undamped collective modes can also propagate in region $4A$, where $\omega_\pm(q) < v_{\rm F} q$.

A better insight into the dispersions of collective modes and their Landau damping inside the EHC could be gained from the 
imaginary parts of the inverse dielectric functions
\be
\Im m\,\left[\frac{1}{\epsilon^{\rm RPA}_{\pm}(q,\omega)}\right]=\Im m\,\left[\frac{1}{1-\lambda_{\pm}(q,\omega)}\right],
\ee
which represent the spectral weights of the collective modes.
The imaginary parts of the inverse dielectric functions have Dirac delta form at $\omega=\omega_\pm(q)$ outside the EHC and acquire finite width inside the continuum due to the Landau damping of the collective modes~\cite{Vignale_Book}. 

\begin{figure}[h]
\centering
\begin{tabular}{c}
\includegraphics[width=\linewidth]{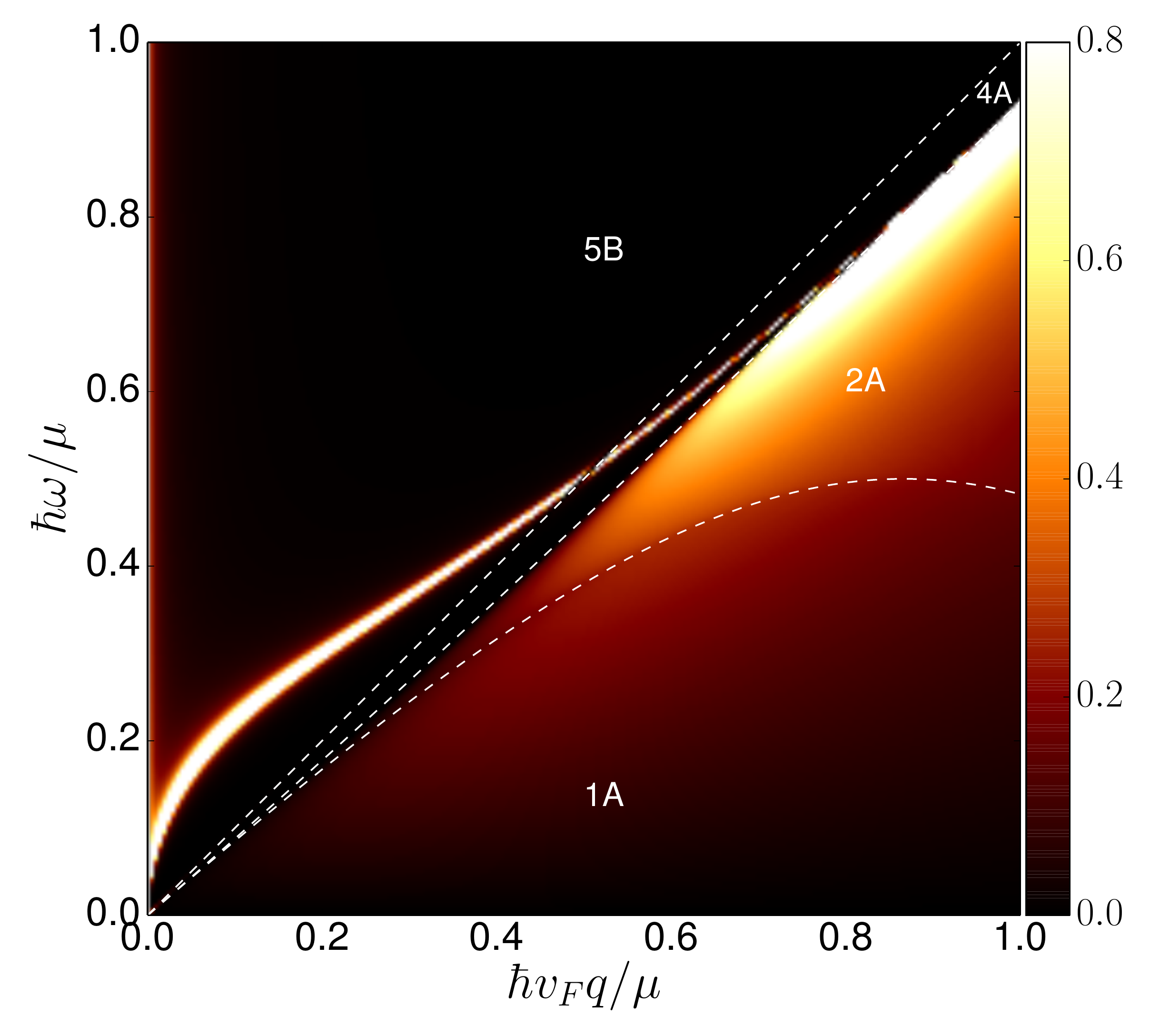}\\
\includegraphics[width=\linewidth]{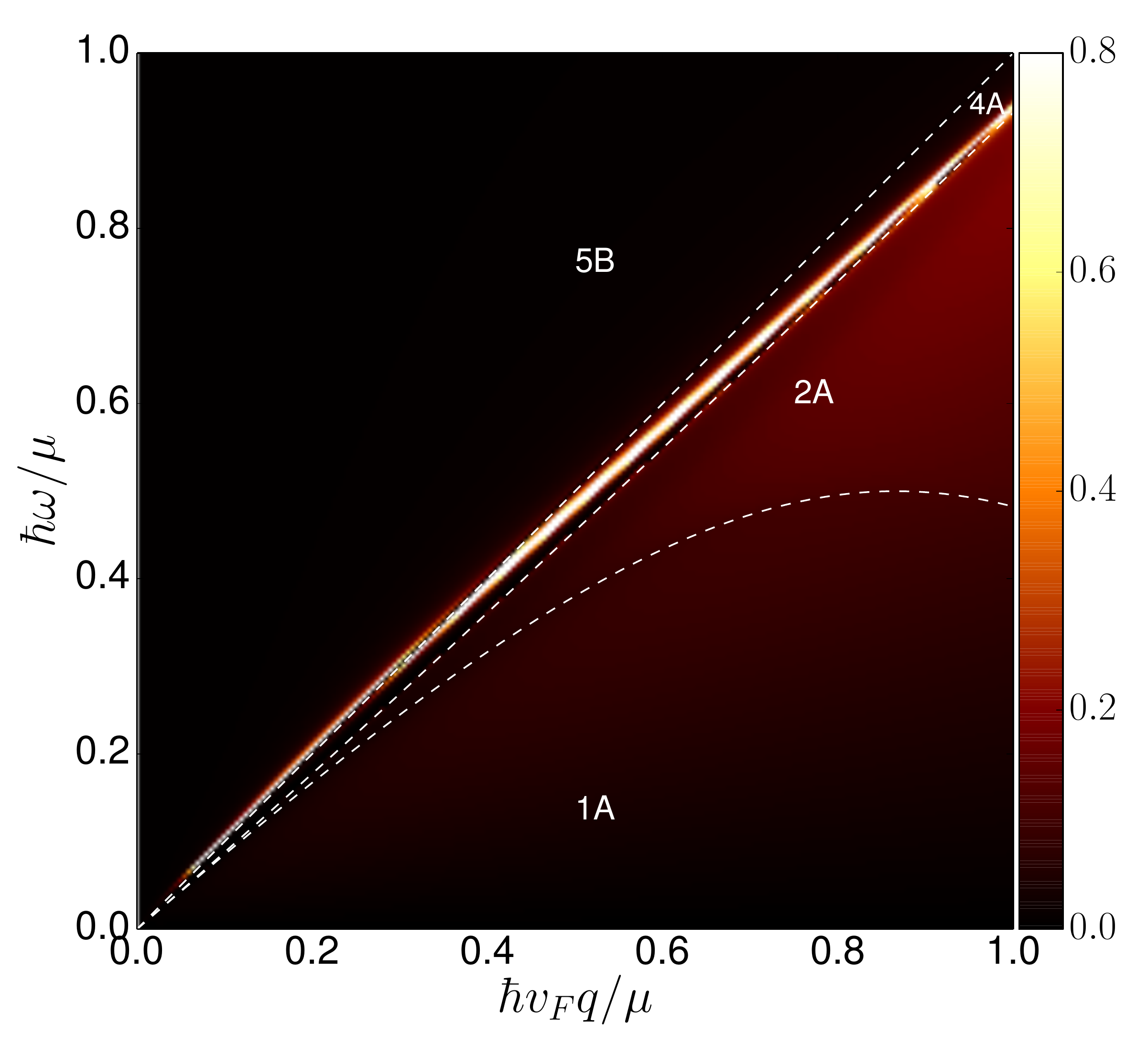}
\end{tabular}
\caption{Top: density plot of the symmetric component of the inverse dielectric function $-\Im m [1/\varepsilon_{+}^{\rm RPA}(q,\omega)]$ versus wave vector $q$ and frequency $\omega$.
In producing this plot we have used $\mu=252$ meV, $t=126$ meV and $\epsilon_{\rm TI}=30$, $\epsilon_{\rm T}=1$, and $\epsilon_{\rm B}=10$ for the dielectric constants of different regions. The Dirac delta peak outside the EHC has been broadened by $ 7\times 10^{-3}$ for better visibility.
Dashed lines show the boundaries between different regions in the $(q,\omega)$-plane as introduced in Appendix~\ref{app:TITF_Response}.
Bottom: same as the top panel, but for the asymmetric components of the inverse dielectric function $-\Im m [1/\varepsilon_{-}^{\rm RPA}(q,\omega)]$, with $\epsilon_{\rm TI}=2$ and the broadening factor of $ 3\times 10^{-2}$.
\label{fig:EEL_plot}}
\end{figure}

Imaginary parts of the symmetric and asymmetric components of the inverse dielectric functions are illustrated in Fig.~\ref{fig:EEL_plot}. The dispersions of collective modes as well as their broadening due to Landau damping inside the EHC are easily recognizable. 

\section{\label{Conclusions}Summary}
We have obtained analytic expressions for the surface resolved dynamical linear density-density response functions of a topological insulator ultra-thin film with finite tunneling between its top and bottom surfaces. 
The full dispersions of the collective density modes and their analytic form at the long-wavelength limit are also investigated.
The velocity of acoustic mode and the Drude weight of the optical mode is tunable through different system parameters. 
The energy of both optical and acoustic modes are suppressed due to the inter-surface tunneling. This means that for a given energy, the collective mode is shifted to large wave-vectors, making them more confined in space. Moreover, the tunneling induced finite gap in the spectrum of surface states makes it possible to switch on or off the plasmon modes through the gate voltage which tunes the chemical potential. Similar behavior has been observed in tunnel coupled double-layer graphene~\cite{Fei2015}. 
These features are expected to be useful for practical implementation in future plasmonic circuitry.

Finally, we note that topological insulator thin films are very interesting systems for the study of different many-body phenomena such as the Coulomb drag effect~\cite{liu_physicaE2016} and excitonic condensation~\cite{moon_epl2012}. Finite tunneling between the two surfaces of a TITF reduces the lifetime of excitons and also makes the detection of Drag current very difficult. However, a double-layer structure consisting of two tunnel coupled topological insulator thin films separated by a thin semiconducting layer which prevents tunneling between two films might be an interesting setup for the investigation of such many-body effects.
 
 \acknowledgments{M.M. is grateful to P.K. Pyatkovskiy for useful comments. S.H.A. is supported by Iran Science Elites Federation (ISEF).}

\appendix

\section{\label{app:TITF_Response} Analytic results for the non-interacting density-density response function of TITF}

\begin{figure}[h]
\centering
\includegraphics[width=\linewidth]{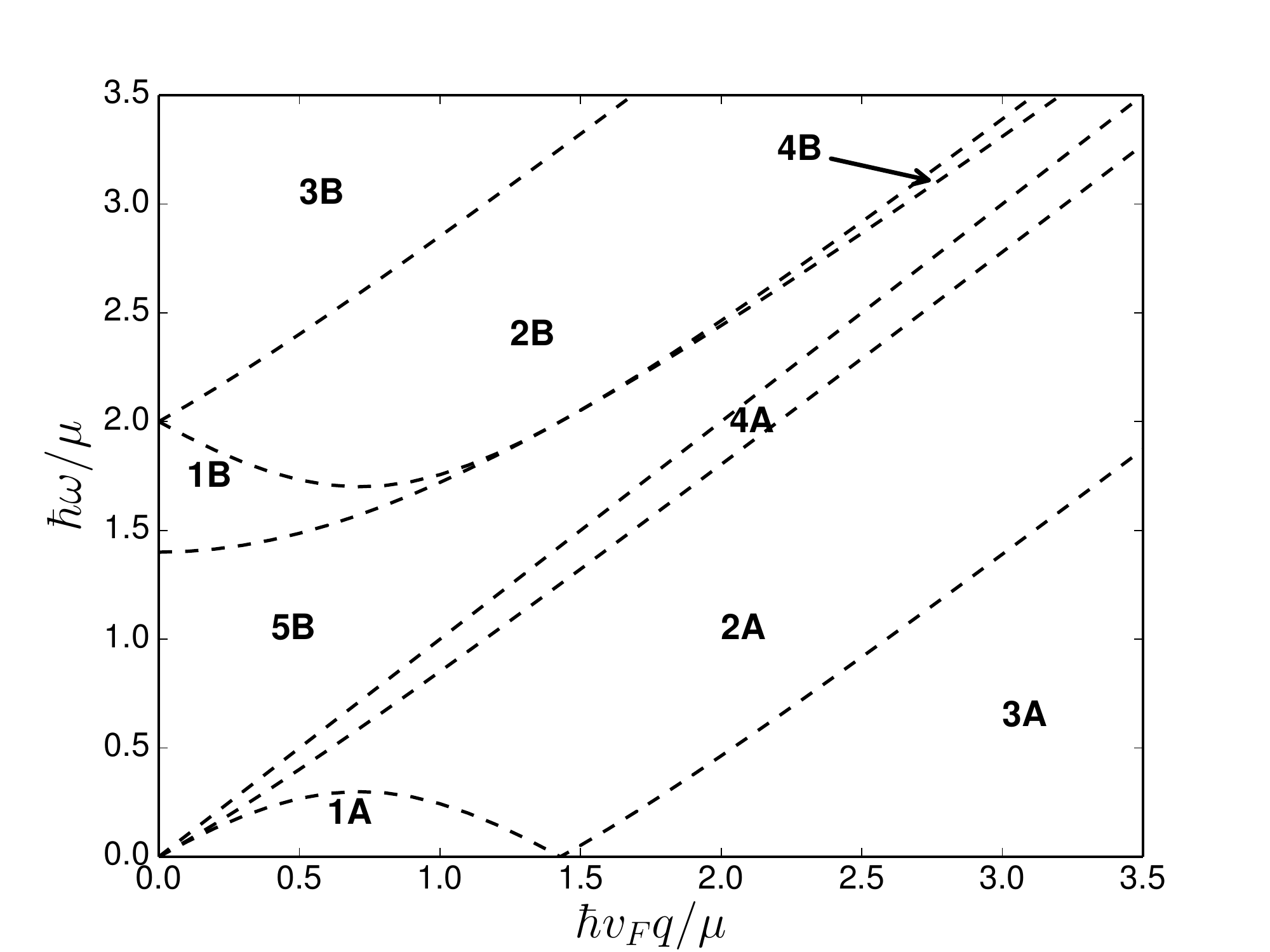}
\caption{Regions with different expressions for the density-density response function. Here we have used $t=0.7 \mu$. 
\label{fig:SPE}}
\end{figure}

In this appendix we calculate the surface resolved non-interacting density-density response functions of a topological insulator ultra-thin film as defined through Eq. \eqref{chi_0}. 
As the total density-density response function $\Pi(q,\omega)=\sum_{l,l'}\Pi_{ll'}(q,\omega)$ of a TITF is identical to the density-density response function of 2D massive Dirac fermions, whose full analytic expressions are available in the literature~\cite{Pyatkovskiy, AThakur2017}, here we will simply present the analytic expressions for the inter-surface density-density response function $\Pi_{12}(q,\omega)$. The intra-surface components could be readily obtained using $\Pi_{11}(q,\omega)=\Pi(q,\omega)/2-\Pi_{12}(q,\omega)$.
From Eqs. \eqref{chi_0} and~\eqref{f12} we find
\be
\Pi_{12}(q,\omega)=
\frac{1}{2S}\sum_{\kv,\lambda,\lambda'}\left( \frac{\lambda \lambda' t^2}{\varepsilon_k \varepsilon_{k'}}\right)\frac{f(\varepsilon_{k\lambda})-f(\varepsilon_{k'\lambda'})}{\hbar\omega+\varepsilon_{k,\lambda}-\varepsilon_{k',\lambda'}+i0^{+}},
\ee
where ${\bf k}' \equiv {\bf k}+{\bf q}$. Assuming $\mu>t$ and $\omega>0$, and following similar procedures as Refs.~\cite{AThakur2017,Pyatkovskiy}, at the zero temperature we find
\be\label{Re_chi_12}
\begin{split}
&\Re e\, \Pi_{12}(q,\omega)=f(q,\omega) x_{1}^{2}\\
&\times\left\{
\begin{array}{ll}
0\, &{\rm1A}\\
\arccos\left(\frac{2\mu-\hbar\omega}{\hbar v_{\rm F} q x_{0}}\right) & {\rm2A} \\
 \arccos\left(\frac{2\mu+\hbar\omega}{\hbar v_{\rm F} q x_{0}}\right)+ \arccos\left(\frac{2\mu-\hbar\omega}{\hbar v_{\rm F} q x_{0}}\right) & {\rm3A} \\
 \arccos\left(\frac{2\mu-\hbar\omega}{\hbar v_{\rm F} q x_{0}}\right)-\arccos\left(\frac{2\mu+\hbar\omega}{\hbar v_{\rm F} q x_{0}}\right)  & {\rm4A} \\
{\,\rm arccosh}\left(\frac{2\mu+\hbar\omega}{\hbar v_{\rm F} q x_{0}}\right)- {\,\rm arccosh}\left(\frac{2\mu-\hbar\omega}{\hbar v_{\rm F} q x_{0}}\right)  &{\rm1B} \\
{\,\rm arccosh}\left(\frac{2\mu+\hbar\omega}{\hbar v_{\rm F} q x_{0}}\right) &{\rm2B} \\
{\,\rm arccosh}\left(\frac{2\mu+\hbar\omega}{\hbar v_{\rm F} q x_{0}}\right)-{\,\rm arccosh}\left(\frac{\hbar\omega-2\mu}{\hbar v_{\rm F} q x_{0}}\right) &{\rm3B} \\
{\,\rm arccosh}\left(\frac{\hbar\omega-2\mu}{\hbar v_{\rm F} q x_{0}}\right)+{\,\rm arccosh}\left(\frac{2\mu+\hbar\omega}{\hbar v_{\rm F} q x_{0}}\right) &{\rm4B} \\
 {\,\rm arcsinh}\left(\frac{2\mu+\hbar\omega}{\hbar v_{\rm F} q \sqrt{-x_{0}^{2}}}\right)-{\,\rm arcsinh}\left(\frac{2\mu-\hbar\omega}{\hbar v_{\rm F} q \sqrt{-x_{0}^{2}}}\right) &{\rm5B} \\
\end{array}
\right.,
\end{split}
\ee
and
\be\label{Im_chi_12}
\begin{split}
&\Im m \,\Pi_{12}(q,\omega)=f(q,\omega) x_{1}^{2}\\
&\times\left\{
\begin{array}{ll}
 {\,\rm arccosh}\left(\frac{2\mu-\hbar\omega}{\hbar v_{\rm F} q x_{0}}\right)- {\,\rm arccosh}\left(\frac{2\mu+\hbar\omega}{\hbar v_{\rm F} q x_{0}}\right) & {\rm1A} \\
- {\,\rm arccosh}\left(\frac{2\mu+\hbar\omega}{\hbar v_{\rm F} q x_{0}}\right) & {\rm2A}\\
0 & {\rm3A}\\
0 & {\rm4A}\\
0 & {\rm1B}\\
 \arccos\left(\frac{2\mu-\hbar\omega}{\hbar v_{\rm F} q x_{0}}\right) &{\rm2B} \\
\pi & {\rm3B} \\
\pi  & {\rm4B} \\
0 & {\rm5B}
\end{array}
\right.,
\end{split}
\ee
for the real and imaginary parts of the inter-surface density-density response function, respectively. 
Here, we have defined $x_{0}= \sqrt{1+4t^2/(\hbar^2 v^{2}_{\rm F}q^2-\hbar^2\omega^2)}$, $x_{1} = 2 t/(\hbar v_{\rm F} q)$, and
\be
 f(q,\omega)=\frac{q^2}{16\pi\sqrt{\left\vert \hbar^2 v^{2}_{\rm F}\,q^2-\hbar^2\omega^2\right\vert}}.
\ee
Different regions in the $(q,\omega)$-plane are introduced according to the arguments of the Dirac delta-functions in the imaginary part of the density-density response function~\cite{Pyatkovskiy, AThakur2017}
\be\label{Regions}
\begin{split}
\left\{
\begin{array}{ll}
{\rm1A:}&\hbar\omega<\mu-\sqrt{\hbar^{2} v^{2}_{F}(q-k_{\rm F})^2+t^2} \\
{\rm2A:}&\pm\mu\mp\sqrt{\hbar^{2} v^{2}_{\rm F}(q-k_F)^2+t^2}<\hbar\omega \\
&<-\mu+\sqrt{\hbar^{2} v^{2}_{\rm F}(q+k_{\rm F})^2+t^2} \\
{\rm3A:}&\hbar\omega<-\mu+\sqrt{\hbar^{2} v^{2}_{\rm F}(q-k_{\rm  F})^2+t^2} \\
{\rm4A:}&-\mu+\sqrt{\hbar^{2} v^{2}_{\rm F}(q+k_{\rm F})^2+t^2}<\hbar\omega< \hbar v_{\rm F}q \\
{\rm1B:}&q<2k_{\rm F}\,,  \sqrt{\hbar^{2} v^{2}_{\rm F}q^2+4t^2}< \hbar\omega \\
&<\mu+\sqrt{\hbar^{2} v^{2}_{\rm F}(q-k_{\rm F})^2+t^2} \\
{\rm2B:}\;&\mu+\sqrt{\hbar^{2} v^{2}_{\rm F}(q-k_{\rm F})^2+t^2}<\hbar\omega \\
&<\mu+\sqrt{\hbar^{2} v^{2}_{\rm F}(q+k_{\rm F})^2+t^2} \\
{\rm3B:}&\hbar\omega>\mu+\sqrt{\hbar^{2} v^{2}_{\rm F}(q+k_{\rm F})^2+t^2} \\
{\rm4B:}&q>2k_{\rm F}\,, \sqrt{\hbar^{2} v^{2}_{\rm F}q^2+4t^2}<\hbar\omega \\
&<\mu+\sqrt{\hbar^{2} v^{2}_{\rm F}(q-k_{\rm F})^2+t^2} \\
{\rm5B:}& \hbar v_{\rm F}q<\hbar\omega<\sqrt{\hbar^{2} v^{2}_{\rm F}q^2+4t^2}
\end{array}
\right.,
\end{split}
\ee
and are also illustrated in Fig.~\ref{fig:SPE}.

\subsection{Results in $|\mu|<t$ limit}
When the chemical potential lies inside the band gap, only the inter-band transitions contribute to the density-density response function, and we find
\be\label{Re_chi0_12}
\begin{split}
& \Re e\, \Pi_{12}^{0}(q,\omega)
=\frac{t^2}{4\pi \hbar^3 v^2_{\rm F} } \frac{(v_{F}^2 q^2-\omega^2)}{ \vert v_{F}^2 q^2 -\omega^2\vert^{3/2}}\\ 
&\times\Bigg\{ \Theta(v_{F}q-\omega)\arccos\left(\frac{v_{F}^2 q^2-\omega^2-4t^2/\hbar^2}{\omega^2-v_{F}^2q^2-4t^2/\hbar^2}\right)\\
&~~~~~~~~~-\Theta(\omega-v_{F}q)\ln\left(\frac{(2t/\hbar+\sqrt{\omega^2-v_{F}^2 q^2})^2}{\vert \omega^2-v_{F}^2q^2-4t^2/\hbar^2\vert}\right)\Bigg\},
\end{split}
\ee
and
\be\label{Im_chi0_12}
 \Im m\, \Pi_{12}(q,\omega)
=\frac{ t^2}{4\hbar^3 v_{\rm F}^2 \sqrt{\omega^2-v_{\rm F}^2 q^2} }  \Theta(\hbar^2\omega^2-\hbar^2v_{\rm F}^2 q^2-4t^2),
\ee
respectively for the real and imaginary parts of the inter-surface density-density response functions.

\subsection{The long-wavelength behaviors}
To obtain the long-wavelength behavior of the collective modes, we investigate the vanishing $q$ limit of the intra-surface and inter-surface density-density response functions, in the dynamical and acoustic limits~\cite{Vignale_Book}, from the full analytic results we have just presented for the $\mu>t$ regime. 

The long-wavelength behavior of the optical mode is obtained from the long-wavelength behaviors of $\Pi_{ll'}(q ,\omega)$ taken in the dynamical limit.
For $\mu>t$, taking the $q\to 0$ and $\omega \to 0$ limits such that $\omega^2/q=c={\rm constant}$, one finds
\be\label{chi_+11_q_go_zero}
\Pi_{11}(q,\omega ) \approx \frac{\mu}{8 \pi \hbar^2 } (1-{\bar t}^2)(2-{\bar t}^2)\frac{ q}{c},
\ee
and
\be\label{chi_+12_q_go_zero}
\Pi_{12}(q,\omega ) \approx \frac{\mu}{8 \pi \hbar^2 } (1-{\bar t}^2){\bar t}^2 \frac{ q}{c},
\ee
where ${\bar t}=t/\mu$.
To obtain the long wavelength behavior of the acoustic mode, the long wavelength behavior of $\Pi_{ll'}(q,\omega)$ in the acoustic (i.e., $\omega/q=y={\rm constant}$) limit 
\be\label{chi_-11_q_go_zero}
\Pi_{11}(q \to  0,y) \approx \frac{\mu}{4\pi  \hbar^2 v^{2}_{\rm F}}
\left[\frac{ y(2-{\bar t}^2)}{\sqrt{y^2 - v^{2}_{\rm F}(1-{\bar t}^2)}}-2\right],
\ee
 and 
 \be\label{chi_-12_q_go_zero}
\Pi_{12}(q \to  0,y) \approx \frac{\mu}{4\pi  \hbar^2 v^{2}_{\rm F}}
\left[\frac{ y{\bar t}^2}{\sqrt{y^2 - v^{2}_{\rm F}(1-{\bar t}^2)}}\right],
\ee
are required.

The elements of the Coulomb interaction matrix Eq.~\eqref{eq:V-matrix} in the long wavelength limit read
\be\label{V_11_q_go_to_zero}
V_{11}(q \to  0)\approx \frac{4\pi e^2}{(\epsilon_{\rm T}+\epsilon_{\rm B})q}-\frac{4\pi e^2 d~(\epsilon_{\rm TI}^{2}-\epsilon_{\rm B}^2)}{\epsilon_{\rm TI}(\epsilon_{\rm T}+\epsilon_{\rm B})^2},
\ee
and
\be\label{V_12_q_go_to_zero}
V_{12}(q \to  0)\approx \frac{4\pi e^2}{(\epsilon_{\rm T}+\epsilon_{\rm B})q}-\frac{4\pi e^2 d~(\epsilon_{\rm TI}^{2}+\epsilon_{\rm T}\epsilon_{\rm B})}{\epsilon_{\rm TI}(\epsilon_{\rm T}+\epsilon_{\rm B})^2}.
\ee
The long wavelength behavior of $V_{22}(q)$ could be obtained from Eq.~\eqref{V_11_q_go_to_zero} upon interchanging  $\epsilon_{\rm T} \leftrightarrow \epsilon_{\rm B}$.

\end{document}